\documentclass[aps,prb,eqsecnum,twocolumn,longbibliography,superscriptaddress,10pt]{revtex4-1}

\usepackage{amsfonts,amssymb,dsfont}
\usepackage[sumlimits,intlimits]{amsmath}
\usepackage{graphics}
\usepackage{graphicx}
\usepackage{color}
\usepackage{mathrsfs}
\usepackage{textcomp}
\usepackage{verbatim}
\usepackage{hyperref}    
\usepackage{bm}          
\usepackage{braket}
\usepackage[normalem]{ulem}

\definecolor{darkgreen}{rgb}{0.0, 0.6, 0.13}

\def\bd{{\bf d}}
\def\bk{{\bf k}}

\def\bn{{\bf n}}

\def\bp{{\bf p}}
\def\br{{\bf r}}

\def\hb{\hat b}

\def\hb{\hat b}

\begin{document}

\title{Energy-level statistics in strongly disordered systems with power-law hopping}

\author{Paraj Titum}
\affiliation{Joint Quantum Institute, NIST/University of Maryland, College Park, MD 20742, USA}

\author{Victor L. Quito}
\affiliation{National Magnetic Field Laboratory, Tallahassee, FL 32310, USA}


\author{Sergey V. Syzranov}
\affiliation{Physics Department, University of California, Santa Cruz, CA 95064, USA}

\date{\today}

\begin{abstract}
Motivated by neutral excitations in disordered electronic materials and
systems of trapped ultracold particles with long-range interactions, we study energy-level statistics
of quasiparticles with the power-law hopping Hamiltonian $\propto 1/r^\alpha$ in a strong random potential.
In solid-state systems such quasiparticles, which are exemplified by neutral dipolar excitations,
lead to long-range correlations of local observables and may dominate energy transport.
Focussing on the excitations in disordered electronic systems,
we compute the energy-level correlation function $R_2(\omega)$ in a finite system in the limit of sufficiently
strong disorder.
At small energy differences
the correlations exhibit Wigner-Dyson statistics.
In particular, in the limit of very strong disorder
the energy-level correlation 
function is given by 
$R_2(\omega,V)=A_3\frac{\omega}{\omega_V}$
 for small frequencies $\omega\ll\omega_V$ and
$R_2(\omega,V)=1-(\alpha-d)A_{1}\left(\frac{\omega_V}{\omega}\right)^\frac{d}{\alpha}
-A_2\left(\frac{\omega_V}{\omega}\right)^2$
for large frequencies $\omega\gg\omega_V$, where 
$\omega_V\propto V^{-\frac{\alpha}{d}}$ is the characteristic matrix element of excitation hopping
in a system of volume $V$, and $A_1$, $A_2$ and $A_3$ are coefficient of order unity
which depend on the shape of the system.
The energy-level correlation function, which we study, allows for a direct experimental observation,
for example, by measuring the correlations
of the ac conductance of the system at different frequencies.
\end{abstract}

\maketitle

In a strongly disordered electronic system, the properties of
charged excitations, such as electrons and holes, are correlated 
on short length scales of order of the localisation length and decay exponentially
with distance. By contrast, neutral excitations, such as localised electron-hole pairs,
allow for long-range hops via virtual processes of annihilating a neutral excitation at one location
and creating it elsewhere.
It has been demonstrated \cite{FleishmanAnderson}, for example, that dipole excitations in 3D {can hop virtually } 
with the distance dependence $\propto 1/r^3$.
Such power-law hops may lead to long-range correlations between physical observables, such as
ac conductivity, {even when charged excitations remain localised}. While
neutral excitations do not carry charge, they are involved in energy transport and, thus,
may dominate heat conductivity. 
Moreover, if certain neutral excitations are delocalised due to the power-law hops 
they may serve as a bath for other excitations~\cite{Burin:ProbablyFirst,Burin:MBLclaims} and thus lead
to the variable-range hopping of charged excitations. 

The dynamics of neutral excitations, therefore, plays a fundamental role in transport and
phase diagrams of granulated materials, superconductive films in the insulating
state~\cite{GantmakherDolgopolov:review}, systems of defects in insulators~\cite{Neumann:defects} and
other disordered systems,
which has motivated recent studies of conductivity~\cite{AleinerEfetov:dipoles} and 
wavefunctions~\cite{Shlyapnikov_Altshuler2016,Shlyapnikov_Kravtsov2017,AleinerEfetov:dipoles,Yao:LevitovCopyPaste} in
 systems with power-law hopping.
Excitations with a generic power-law hopping $\propto1/r^\alpha$ with tunable $\alpha$ have also been
realised recently in 1D\cite{Monroe:longrange,Islam:longrange,Blatt:chain1,Blatt:chain2} and
2D\cite{Bollinger:longrange} arrays of trapped ultracold ions.
Such systems may be used to simulate disordered electronic materials, yet serve as a platform
for observing novel fundamental phenomena, for example,
many-body-localisation transitions\cite{BAA} or high-dimensional disorder-driven effects\cite{Syzranov:review}. 
In this paper we study analytically the energy-level statistics (ELS) of excitations with power-law hopping Hamiltonians 
in strongly disordered systems.

Energy-level statistics in a disordered system reflects fundamental symmetries 
and {is often} used to diagnose conducting and insulating phases at different
disorder strengths.
Abundant numerical data (see Ref.~\onlinecite{Efetov:book} for a review)
suggests also that the ESL is linked to the chaotic properties of a system; systems such as
chaotic billiards and disordered metals display chaotic or non-chaotic behaviour depending
on whether their statistics is {Wigner}-Dyson or Poisson.  
Recently, ESL has also received much renewed attention
in the context of many-body-localisation transitions\cite{BAA}; interacting disordered systems 
are expected~\cite{HusePal} to display Poisson or Wigner-Dyson statistics of the {\it many-body}
levels of the system in many-body-delocalised and many-body-localised states, respectively\cite{HusePal,Serbyn:LevelStatistics,TikhnonovMirlin:AlphaSimulations} .
Similarly, ESL has been demonstrated numerically\cite{Garcia:SYKtransition} to distinguish between
chaotic and non-chaotic behaviour in a
generalised Sachdev-Ye-Kitaev model\cite{SachdevYe}, which is often used
as a toy model in the studies of quantum chaos.
In such studies the many-body ELS at the transition is used 
as a
numerical tool for detecting a phase transition, although, unlike the single-(quasi)particle
level statistics, cannot be sraightforwardly measured in condensed-matter experiments.
It has also been conjectured recently\cite{Syzranov:ChaosTransition} that 
the ELS of single quasiparticles reflects an interaction-driven transition between chaotic and non-chaotic behaviour.

{The relation between phase transitions and excitation statistics can be explored further by  analysing the ELS in the respective phases.} Although the ELS of neutral excitations
in insulating materials also determines the heat transport and correlations between local ac responses, e.g.
the ac conductivity, it has largely avoided researchers' attention, in contrast with the statistics
of charged excitations\cite{Efetov:book,AltshulerShklovskii,Burmistrov:MottStatistics}.
In this paper,
focussing on neutral excitations in disordered solids ($\alpha=3$),
 we compute microscopically the correlation functions of energy levels 
of excitations with the power-law hopping $\propto 1/r^\alpha$.


This paper is organised as follows. Our main results for the energy-level correlation functions
are summarised in Sec.~\ref{Sec:Summary}. In Sec.~\ref{Sec:Model} we discuss the model of dipole
excitations, the simplest type of neutral excitations in a disordered system. 
Sec.~\ref{Sec:StrongDis} deals with the statistics of dipole excitations in strongly disordered systems
in dimensions $d<3$ and sufficiently small disordered 3D systems. The case of a 3D system, which requires
a special consideration, is addressed in Sec.~\ref{Sec:Weaker_Disorder}.
Section~\ref{Sec:GenericPowerLaw} is devoted to the energy-level correlation functions in systems with a
power-law hopping $\propto 1/r^\alpha$ with an arbitrary power $\alpha$, which have recently been realised in experiments
with trapped ultracold particles.
We conclude in Sec.~\ref{Sec:conclusion}.


\section{Summary of results}

\label{Sec:Summary}

In this paper, we characterise the statistics of the
energy levels of excitation in a disordered system of a finite volume $V$ by the correlation function
\begin{equation}
 R_{2}\left(\omega\right)=
 {\left\langle \nu\left(E+\frac{\omega}{2}\right)\nu\left(E-\frac{\omega}{2}\right)\right\rangle_{\text{dis}} }/
 {\left\langle \nu\left(E\right)\right\rangle_{\text{dis}} ^{2}},
 \label{R2definiiton}
\end{equation}
where $\nu(E)$ is the density of states (DoS) of the excitations
for a specific disorder realisation, and $\langle\ldots\rangle_{\text{dis}}$
is our convention for disorder averaging. 

In the limit of very strong disorder, spatial and energy correlations between observables in electronic systems
are dominated by dipole excitations, i.e. pairs of electron and hole excitations located close to each other. 
Such dipoles allow for long-range hops with the $\propto1/r^3$ distance dependence\cite{FleishmanAnderson}.
In principle, the hopping of more complicated excitations, consisting of multiple
electrons and holes, has in general the same distance dependence, but is suppressed due to smaller matrix elements of
recombination of those excitations at strong disorder.

The density of states of
dipole excitations is proportional to the ac conductance~\cite{Syzranov:ac} of the system. 
This allows one to observe the correlation function (\ref{R2definiiton}) in experiment, e.g.,
by measuring the correlations of ac conductance $G(\tilde\omega)$ of the system as a function
of frequency $\tilde\omega$
and computing the correlator
$R_2 (\omega) \propto \left<G(\Omega)G(\Omega+\omega)\right>_\Omega$,
where $\langle\ldots\rangle_\Omega$ is the averaging with respect to the frequency $\Omega$
in a sufficiently large interval of energies. 

{\it Very strong disorder in an electronic system.} In the case of very strong disorder in a solid-state
system, we find
\begin{align}
	R_{2}(\omega,V)\approx
	\left\{
	 \begin{array}{cc}
	 1-(3-d)C_{1}\left(\frac{\omega_V}{\omega}\right)^\frac{d}{3}
	 -C_{2}\left(\frac{\omega_V}{\omega}\right)^2 , & \omega \gg\omega_V
	 \\
	  C_3 \frac{\omega}{\omega_V} , & \omega \ll\omega_V
	 \end{array}
	\right.
	\label{R2MainResult}
\end{align}
where the characteristic frequency $\omega_V$ scales with volume the volume $V$ of the system 
as $\omega_V\propto V^{-\frac{3}{d}}$; and the coefficients 
$C_1$, $C_2$ and $C_3$ are independent 
of the volume $V$, but the coefficients $C_2$ and $C_3$ 
depend on the shape of the system.
This result applies to all strongly disordered systems in dimensions $d<3$
and to sufficiently small 3D systems. In such systems, the correlations between energy levels come from
rare resonances between pairs of excitation states which are located far from each other but have close energies. 

{\it 3D systems.} In 3D electronic systems, unlike the case of lower dimensions, excitation states involve 
resonances on multiple sites\cite{Levitov_1990}, which is why the 3D case requires a special consideration.
We find that in 3D systems the energy-level correlation function $R_2(\omega)$ 
is still given by Eq.~(\ref{R2MainResult}) with the coefficients $C_i$ independent of volume $V$ only
in the limits of small and large volumes. However, for a 3D system of arbitrary size, these coefficients
have an explicit $V$ dependency, and, thus, the scaling of the correlation function with system size is
different.

{\it Arbitrary power-law hopping.} The results for the energy-level correlations in a system of 
neutral electronic excitations
with the $\propto 1/r^3$ hopping may be generalised to the case of an arbitrary power-law hopping $\propto1/r^\alpha$.
Such hopping with arbitrary $\alpha$ my be realised, for example, in arrays of trapped ultracold
ions\cite{Monroe:longrange,Islam:longrange,Blatt:chain1,Blatt:chain2,Bollinger:longrange}
in optical or magnetic traps. For such hopping we obtain
\begin{align}
	R_{2}(\omega,V)\approx
	\left\{
	 \begin{array}{cc}
	 1-(\alpha-d)A_{1}\left(\frac{\omega_V}{\omega}\right)^\frac{d}{\alpha}
	 -A_2\left(\frac{\omega_V}{\omega}\right)^2 , & \omega \gg\omega_V
	 \\
	 A_3\frac{\omega}{\omega_V} , & \omega \ll\omega_V
	 \end{array}
	\right.,
\end{align}
where $\omega_V\propto V^{-\frac{\alpha}{d}}$. 
The power-law dependency of the correlation function $R_2$ on the energy difference $\omega$ may signal
of a possible chaotic behaviour of the dynamics of the excitations, as we discuss in Sec.~\ref{Sec:conclusion}.


\section{Model for neutral excitations in solids}

\label{Sec:Model}

In what immediately follows we describe the effective Hamiltonian of electron-hole dipoles, 
the simplest type of neutral excitations in an electronic system.
The Hamiltonian of the dipole excitations in a disordered medium is given by 
\begin{eqnarray}
&&{\mathcal{\hat H}_{0}}  =
\sum_{\br,\bd}E_{\br\bd}
\hb_{\br\bd}^{\dagger}\hb_{\br\bd}
-\sum_{\br,\bd}J_{\br\bd}\left(\hb_{\br\bd}^{\dagger}+\hb_{\br\bd}\right)
\nonumber\\
 &&+\sum_{\boldsymbol{rd,r'd'}}E_{\text{int}}\left(\br\bd,\br'\bd'\right)
 \hb_{\br\bd}^{\dagger}\hb_{\br'\bd'}^{\dagger}\hb_{\br'\bd'}
 \hb_{\br\bd},
\end{eqnarray}
where $\hb^{(\dagger)}_{\br\bd}$ is the annihilation (creation) operator of a dipole 
with polarisation $\bd$ at location
$\br$ (e.g., the location
of the positive charge in the dipole);
$E_{\br\bd}$ is the energy of the respective dipole state, which strongly fluctuates
from site to site due to the presence of quenched disorder; $J_{\br\bd}$ is the matrix element
of the recombination of the dipole (electron hopping to the location of the hole or vice a versa),
which may be assumed real without loss of generality;
we have also introduced the interaction energy
\begin{align}
	E_{\text{int}}\left(\br\bd,\br'\bd'\right)
	=Q(\hat{\bd},\hat{\bd^{\prime}},\boldsymbol{n})\frac{|\bd||\bd^\prime|}{|\br-\boldsymbol{r'}|^3},
\end{align}
\begin{align}
	Q(\hat{\bd},\hat{\bd^{\prime}},\boldsymbol{n})=
	\hat{\bd}\cdot\hat{\bd^{\prime}}-3(\hat{\bd}\cdot\boldsymbol{n})(\hat{\bd^{\prime}}\cdot\boldsymbol{n})
	\label{QDefinition}
\end{align}
between dipoles at locations $\br$ and $\br^\prime$ with polarisations $\bd$ and $\bd^\prime$, respectively,
where $\bn=({\br-\br^\prime})/{|\br-\br^\prime|}$ and
$\hat{\bd}=\bd/|\bd|$ are unit vectors parallel, respectively, to $\br-\br^\prime$ and $\bd$.

{\it Long-range hopping.} While strong disorder prevents dipole hopping on short distances,
dipole excitations allow for long-range virtual hops\cite{FleishmanAnderson} between remote sites with close energies
via annihilating a dipole at the initial location and
then creating it at the final location or vice a versa.
The amplitude of such a hop is given, to the leading order in the recombination
elements $J_{\br\bd}$, by
\begin{equation}
 T_{\br\bd,\br'\bd'}
 \approx \frac{J_{\br\bd}J_{\br^\prime \bd^\prime}}{E_{\br\bd}^{2}}\frac{|\bd||\bd^\prime|Q(\hat{\bd},\hat{\bd^{\prime}},\boldsymbol{n})}{\left|{\bf r-{\bf r}'}\right|^3}.
 \label{Tunnelling}
\end{equation}
Thus, in the limit of a strongly disordered system (small recombination elements $J_{\br\bd}$ compared
to the typical fluctuations of the energies $E_{\br\bd}$), the dynamics of the dipoles is effectively a single-particle
problem.

{\it Cotennelling through excitation states with high energies.}
When computing energy-level correlations,
we assume that the energies of 
all sites $E_{\br\bd}$ lie sufficiently close to each other
and do not consider sites with high excitation energies $E_{\br\bd}\gg\omega$.
In principle, cotunnelling through such high-energy sites in a disordered system
may lead to ultraviolet divergencies
in physical observables~\cite{Syzranov:RGbosons}. These divergencies may be treated by 
means of the renormalisation-group (RG) procedure described in Ref.~\onlinecite{Syzranov:RGbosons}, which
repeatedly removes highest-energy excitation states from the system
while renormalising transition amplitudes between all other states, with energies closer to $\omega$.
In this paper, we assume that the system we consider 
is already renormalised following this procedure and the recombination elements $J_{\br\bd}$
already include virtual cotunnelling through excitation states with energies far from $E$.


\section{Strong disorder in solids}
\label{Sec:StrongDis}

In this section we consider the correlations of dipole energy levels and ac conductances
in a strongly disordered system, in which most dipole states are localised almost entirely on single sites
$(\br,\bd)$ (with given polarisations) and are weakly perturbed by the tunnelling to other sites.
A dipole is localised almost entirely on one site 
$(\br,\bd)$ if there are no other ``resonant'' sites around it\cite{FleishmanAnderson,Levitov_1990,Levitov:AnnReview,Levitov2}
with close energies, 
$|E_{\br\bd}-E_{\br^\prime \bd^\prime}|\lesssim |T_{\br\bd,\br^\prime\bd^\prime}|$.

The number of resonant sited around a given site $\br$ may be estimated as
$N_{\br\bd}\sim\nu_0 n\sum_{\bd^\prime}
\int_{\br^\prime} \left|T_{\br\bd,\br^\prime\bd^\prime}\right|d{\br^\prime}$,
where $n$ is the concentration of sites and $\nu_0$ is the density of dipole states at an isolated site.
For the hopping element $|T_{\br\bd,\br^\prime\bd^\prime}|\propto 1/|\br-\br^\prime|^3$, 
given by Eq.~(\ref{Tunnelling}) and for strong disorder, the average number of resonant sites is significantly
smaller than unity near each given site in dimensions $d<3$. In this regime, 
most of the dipoles are strongly localised and their states may be considered
unaffected by the resonant sites. 
In 3D, however, the number of resonant sites
diverges\cite{FleishmanAnderson,Levitov_1990,Levitov:AnnReview,Levitov2} $\propto 
n\nu_0\langle \bd^2\rangle J^2 E_{\br\bd}^{-2}\ln(L/a)$ in the limit of an infinite
system, $L\rightarrow\infty$, where $a$ is the typical size of a dipole excitation
and $J$ is the characteristic recombination matrix element. 

In this section, we assume that the disorder is strong and that
most dipoles are strongly localised,
either due to a low spatial dimension $d<3$ or due to a sufficiently small size in 3D,
$L\ll\exp\left(\frac{E^2}{\nu J^2 d^2}\right)$.

\subsection{Generic expressions for dipolar energy-level correlations}\label{subsec:Strong_disorder-Energy_level_correlations}

In what immediately follows we derive a generic expression for the correlator $R_2(\omega,\br,\br^\prime)$
of the energy levels
of a dipole excitation on two sites and then, using it, compute the correlator $R_2(\omega)$ of the energy levels
in a strongly disordered system of volume $V$.

While most dipoles are strongly localised on single sites,
there exist rare pairs of sites with close energies, on which dipole states get strongly 
hybridised due to the tunnelling.
This hybridisation of pairs of sites
leads to correlations between dipole states on arbitrarily long distances,
which lead to correlations between energy levels and various observables. 
In the limit of strong disorder (small system size) under consideration, one may neglect
resonances between clusters of three or more sites.

{\it Correlation function in a system of two sites.}
The hybridisation of two dipole states with close energies $E_{\br\bd}$ and $E_{\boldsymbol{r'd'}}$
at locations
$\br$ and $\br^\prime$
and with polarisations $\bd$ and  $\bd^\prime$ leads to the creation of two hybridised states with
energies
\begin{align}
E_\pm =\frac{1}{2}\left(E_{\boldsymbol{r'd'}}+E_{\br\bd}\right)
\pm\frac{1}{2}\left[(E_{\br\bd}-E_{\br^\prime\bd^\prime})^2+4\left|T_{\br\bd,\br'\bd'}\right|^2\right]^\frac{1}{2},\label{eq:Epm}
\end{align}
where the
hopping amplitude $T_{\br\bd,\br'\bd'}$ 
is given by Eq.~\ref{Tunnelling}.
The density
of states on such a pair of sites is
given by $\nu(E)=\delta\left(E-E_{+}\right)+\delta\left(E-E_{-}\right)$.
The 
correlation function $R_2(\omega,\br,\br^\prime)$ of dipole states on two sites $\br$
and $\br^\prime$ is given by

\begin{align}
R_{2}\left(\omega,\br,\br^\prime\right)=&\frac{1}{4\nu_0^2}\int d\bd_{1} d\bd_{2}\ f\left(\bd_{1}\right)f\left(\bd_{2}\right)
\nonumber\\
&
\left<
\left[\delta\left(E+\frac{\omega}{2}-E_{+}\right)+\delta\left(E+\frac{\omega}{2}-E_{-}\right)\right]
\right.
\nonumber\\
& \left.\left[\delta\left(E-\frac{\omega}{2}-E_{+}\right)+\delta\left(E-\frac{\omega}{2}-E_{-}\right)\right]
\right>_{\text{dis}}
\label{R2initial}
\end{align}
where $f(\bd)$ is the distribution function of the dipole moments, assumed independent of 
the on-site energy fluctuations;
$\langle\ldots\rangle_{\text{dis}}$ is the averaging with respect to the realisations of disorder,
which affects both the energies $E_{\br\bd}$ and $E_{\br^\prime d^\prime}$ and the hopping
$T_{\br\bd \br^\prime d^\prime}$ via the recombination elements $J_{\br\bd}$;
we have also used that the density of dipole states $\nu_0$ at each site may be considered 
constant close to the energy $E$ under considerations, so long as $\omega\ll E, \nu_0^{-1}$.

To make further progress, we assume that the recombination elements $J_{\br\bd}$ and the on-site 
energies $E_{\br\bd}$ fluctuate independently. Introducing a variable $\tau$, such that $2T_{\br\bd,\br'\bd'}\tau= E_{\br\bd}-E_{\boldsymbol{r'd'}}$, and
the distribution function $P(J_{\br\bd})$ of the recombination elements $J_{\br\bd}$, the two-site 
correlator (\ref{R2initial}) is reduced to
\begin{align}
R_{2}\left(\omega,\br,\br^\prime\right)=\frac{1}{2}\int dJ_{\br\bd}\ dJ_{\br^\prime\bd^\prime}
\ P(J_{\br\bd})P(J_{\br^\prime\bd^\prime})
\nonumber \\
\int d\bd_{1}d\bd_{2}\ f\left(\bd_{1}\right)f\left(\bd_{2}\right) 
\int d\tau\ \left|T_{\br\bd,\br'\bd'}\right|
\nonumber\\
  \delta\left(\omega-2\left|T_{\br\bd,\br'\bd'}\right|\sqrt{\tau^{2}+1}\right).
  \label{R22}
\end{align}

{\it Correlation function on multiple sites.} Eq.~(\ref{R22})
describes dipole energy correlations in a system of two sites.
In what immediately follows, we derive the energy-level correlation function
for a system of $N\gg1$ sites.

In the absence of dipole tunnelling between sites ($T_{\br\bd,\br^\prime\bd^\prime}=0$), it is given by
\begin{align}
 R_2^{\rm uncorr}(\omega)=
\nonumber\\
 \frac{\sum_{\br\bd,\br^\prime\bd^\prime}\int dE_{\br\bd}dE_{\br^\prime\bd^\prime}
\nu_0^2 \delta(E-E_{\br\bd})\delta(E+\omega-E_{\br^\prime\bd^\prime})}
{(\nu_0 N)^2}
\nonumber
\\
=\frac{N(N-1)}{N^2}\overset{N\rightarrow\infty}{\longrightarrow} 1
\end{align}
For finite tunnelling, dipole states on different sites get hybridised, and the correlation function
$R_2(\omega)$ deviates from unity. In the limit of strong disorder, which we consider in this section,
pairs of resonant sites are rare, only a small fraction of dipole states get hybridised due to dipole
hopping and resonances of three or more sites may be neglected.

The correlation function in such a system of multiple sites may be found by hybridising dipole
states on all pairs of sited with close energies and computing (see Appendix~\ref{Sec:R2modificationStep} for details)
the modification of the correlation function
similarly to Eq.~(\ref{R22}).
The full correlation function in a system of many sites with rare resonances is given by
\begin{widetext}
 \begin{align}
   R_{2}\left(\omega\right)=\frac{1}{V^2}
   \int dJ_{\br\bd}\ dJ_{\br^\prime\bd^\prime}
   \ P(J_{\br\bd})P(J_{\br^\prime\bd^\prime})
   \int d\br d\br^\prime
   \int d{\bd}\ d\bd^\prime\ f\left(\bd\right)f\left(\bd^\prime\right)
   \ \Theta\left[1-\left({2T_{\br\bd,\br'\bd'}}/\omega\right)^{2}\right]
   /\left[{1-\left({2T_{\br\bd,\br'\bd'}}/{\omega}\right)^{2}}\right]^\frac{1}{2}, 
   \label{eq:R_2_Main}
 \end{align}
 \end{widetext}
where the hopping element 
$T_{\br\bd,\br'\bd'}$ is given by Eq.~(\ref{Tunnelling})
and $\Theta(\ldots)$ is the theta-function.

A rigorous evaluation of the correlation function $R_2(\omega)$, given by Eq.~(\ref{eq:R_2_Main}),
requires making an assumption about the distribution $P(J_{\br\bd})$ of the electron-hole recombination matrix
elements $J_{\br\bd}$. Since the exact form of the distribution will affect only numerical coefficient,
we assume, for simplicity, that they are sharply peaked near
certain value $J$, 
\begin{align}
	P(J_{\br\bd})=\delta(J-J_{\br\bd}).
	\label{eq:P(J)}
\end{align}
We will also assume a {uniformly random} orientation of the dipole moments in the d-dimensional
space, 
\begin{align}
f(\bd)=
\delta\left(\left|\bd\right|-d_{0}\right)/\left(\Omega_{d}d_{0}^{d-1}\right),
\end{align}
where $\Omega_d=\frac{2\pi^{\frac{d}{2}}}{\Gamma\left(\frac{d}{2}\right)}$
is the area of a unit sphere in $d$ dimensions.

We note that in general the recombination elements $J_{\br\bd}$ may have arbitrary signs, in contrast
to our choice of their distribution (\ref{eq:P(J)}). 
Indeed, in a system of electrons in a disordered system such elements are determined by
overlap integrals of oscillating wavefunctions. However, such fluctuations of the sign do not
affect qualitatively the correlation function $R_2(\omega)$ in the limit of strong disorder considered 
in this section.

By switching to the integration with respect to the centre-of-mass $\frac{\br+\br^\prime}{2}$ and 
relative $\tilde{\br}=\br-\br'$ coordinates,
Eq.~(\ref{eq:R_2_Main}) may be simplified as
\begin{align}
	R_{2}\left(\omega\right)=&\frac{1}{V}\int d\bd\ d{\bd'}\ f\left(\bd\right)f\left(\bd^\prime\right)
	\nonumber \\
	&\int_{\tilde{r}>\tilde{r}_\omega}d\tilde{\boldsymbol{r}}\left[1-\left(
	\frac{\tilde{r}_\omega(\bd_{1},\bd_{2},\tilde{\boldsymbol{r}}/\tilde{r})}{\tilde{r}}\right)^{6}\right]^{-\frac{1}{2}}
\label{eq:R2_interm1}
\end{align}
where we have introduced $\tilde{r}_\omega\left(\bd_{1},\bd_{2},\boldsymbol{n}\right)=\left[\frac{2J^{2}d_0^2}{E^2}\left(\frac{\left|Q\left(\bd_{1},\bd_{2},\boldsymbol{n}\right)\right|}{\omega}\right)\right]^{\frac{1}{3}}$, the characteristic distance at which the 
the dipole interaction energy is of order $\omega$.
The integral in Eq.~(\ref{eq:R2_interm1}) cannot be evaluated exactly
for arbitrary parameters. Below we compute the asymptotic 
behaviour of the correlation function $R_2(\omega)$ in the limit of
large $\omega\gg\omega_V$ and small $\omega\ll\omega_V$ frequencies, where
\begin{align}
	\omega_V=
	\frac{2 J^{2}d_{0}^{2}}{E^{2}V^{\frac{3}{d}}}
\end{align}
is the characteristic interaction energy between dipoles on the length of the 
system $L\sim V^{-\frac{1}{d}}$. The scale $\omega_V$ plays the role of the characteristic
energy level splitting in a system of volume $V$.

\subsubsection{Large-frequency limit}

For $\omega\gg\omega_V$, the value of the
integral (\ref{eq:R2_interm1}) in dimensions $d<3$ comes from distances $\tilde{r}$
of order $\tilde{r}_\omega$, which in this limit are significantly shorter than the 
characteristic system size $V^\frac{1}{d}$. The upper limit of integration with respect to $\tilde{r}$
in Eq.~(\ref{eq:R2_interm1}) may be extended to infinity, giving
\begin{align}
	R_{2}\left(\omega\gg\omega_V\right)\approx
	1-\frac{2\pi^\frac{d+1}{2}\Gamma\left(1-\frac{d}{6}\right)}
	{\Gamma\left(\frac{d}{2}\right)\Gamma\left(\frac{1}{2}-\frac{d}{6}\right)d}
	\left\langle \left|Q\right|^{\frac{d}{3}}\right\rangle_\bd 	
	\left(\frac{\omega_V}{\omega}\right)^{\frac{d}{3}}
	\label{R2StrongGen}
\end{align}
where $\left\langle \left|Q\right|^{\frac{d}{3}}\right\rangle _{\bd}=\int d\hat{\bd}_{1}d\hat{\bd}_{2}d\boldsymbol{n}\left|Q\left(\hat{\bd}_{1},\hat{\bd}_{2},\boldsymbol{n}\right)\right|^{\frac{d}{3}}$
is our convention
for the function $\left|Q\right|^{\frac{d}{3}}$ averaged with respect to the directions 
$\hat{\bd}_1$ and $\hat{\bd}_2$ of the dipole moments $\bd_1$ and $\bd_2$
and of the vector $\bn$, where $Q$ describes the angular dependence of the dipolar interactions [cf. Eq.~(\ref{QDefinition})].

Equation (\ref{R2StrongGen}) 	 is our main result for the energy-level correlation function in a generic
strongly disordered system in $d$ spatial dimensions in the limit of large frequencies energy differences $\omega$.
For $d=2$ it gives
\begin{equation}
R_{2}^{d=2}\left(\omega\gg\omega_V\right)=1-1.28 \left(\frac{\omega_V}{\omega}\right)^{\frac{2}{3}} \label{eq:d_2_behavior}.
\end{equation}
where we used $\left\langle \left|Q\right|^{\frac{2}{3}}\right\rangle \approx0.95$ in two dimensions.
Let us note that for $d=3$ the coefficient before the last term in Eq.~(\ref{R2StrongGen}) vanishes,
and, in order to obtain the frequency dependency of $R_2(\omega)$, it is necessary to evaluate a correction
of higher order in $1/\omega$. 
Direct integration in Eq.~(\ref{eq:R2_interm1}) yields
\begin{equation}
R_{2}^{d=3}\left(\omega\gg\omega_V\right)=1-C_{2}\left(\frac{\omega_{V}}{\omega}\right)^{2}
\label{eq:R2_large_omega_d=3_strong_disorder},
\end{equation}
where $C_2$ is a coefficient of order unity which depends on the shape of the sample.

Eq.~(\ref{R2StrongGen}), which accurately describes the large-frequency behaviour
of correlation function in dimensions $d\neq 3$, and Eq.~(\ref{eq:R2_large_omega_d=3_strong_disorder}),
which applies for $d=3$,
may be combined into the interpolation formula
\begin{align}
R_{2}\left(\omega\right)\approx1-\left(3-d\right)C_{1}\left(\frac{\omega_{V}}{\omega}\right)^{\frac{d}{3}}-C_{2}\left(\frac{\omega_{V}}{\omega}\right)^{2}.
\label{Interpolation2}
\end{align}
We emphasise that in general in dimensions $d\neq3$ the correlation function $R_2(\omega)$ contains contributions $\propto 1/\omega^\beta$
with $d/3<\beta<2$, which exceed the last term in Eq.~(\ref{Interpolation2}).
However, in these dimensions the leading-order large-frequency behaviour of the energy-level correlations
is determined by the first term in the right-hand side of Eq.~(\ref{Interpolation2}). 
In 3D, the leading order correlations are described by the last term of (\ref{Interpolation2}).
Thus, Eq.~(\ref{Interpolation2}) accurately describes the large-frequency asymptotics of the
of the correlation function $R_2(\omega)$ in all dimensions.

\subsubsection{Small-frequency limit}

In the limit $\omega\ll\omega_V$ Eq.~(\ref{eq:R2_interm1}) may be 
rewritten as
\begin{align}
R_{2}\left(\omega\right)=C_{3}\frac{\omega}{\omega_{V}}
\label{R2omega}
\end{align}
{where the coefficient $C_{3}$ is given by $C_{3}=\frac{\pi}{V^{1+\frac{3}{d}}}\int d\tilde{\boldsymbol{r}}\,\tilde{r}^{3}\int d\bd_{1}d\bd_{2}f\left(\bd_{1}\right)f\left(\bd_{2}\right)\delta\left(Q\right)$.} 

Equation (\ref{R2omega}) is our main result for the dipole energy-level correlation function
in a strongly disordered electronic system in arbitrary dimensions in the limit of small frequencies. 
It demonstrates that the correlation function is linear in frequency $\omega$
for such strongly disordered systems, similarly to the case of a weakly disordered metal in the orthogonal symmetry
class~\cite{Efetov:book}. Such a linear dependency may also be expected from a phenomenological
random-matrix-theory argument\cite{AltshulerShklovskii}, based on considering a two-level system with a random Hamiltonian.

\begin{figure}
\includegraphics[width=0.8\linewidth]{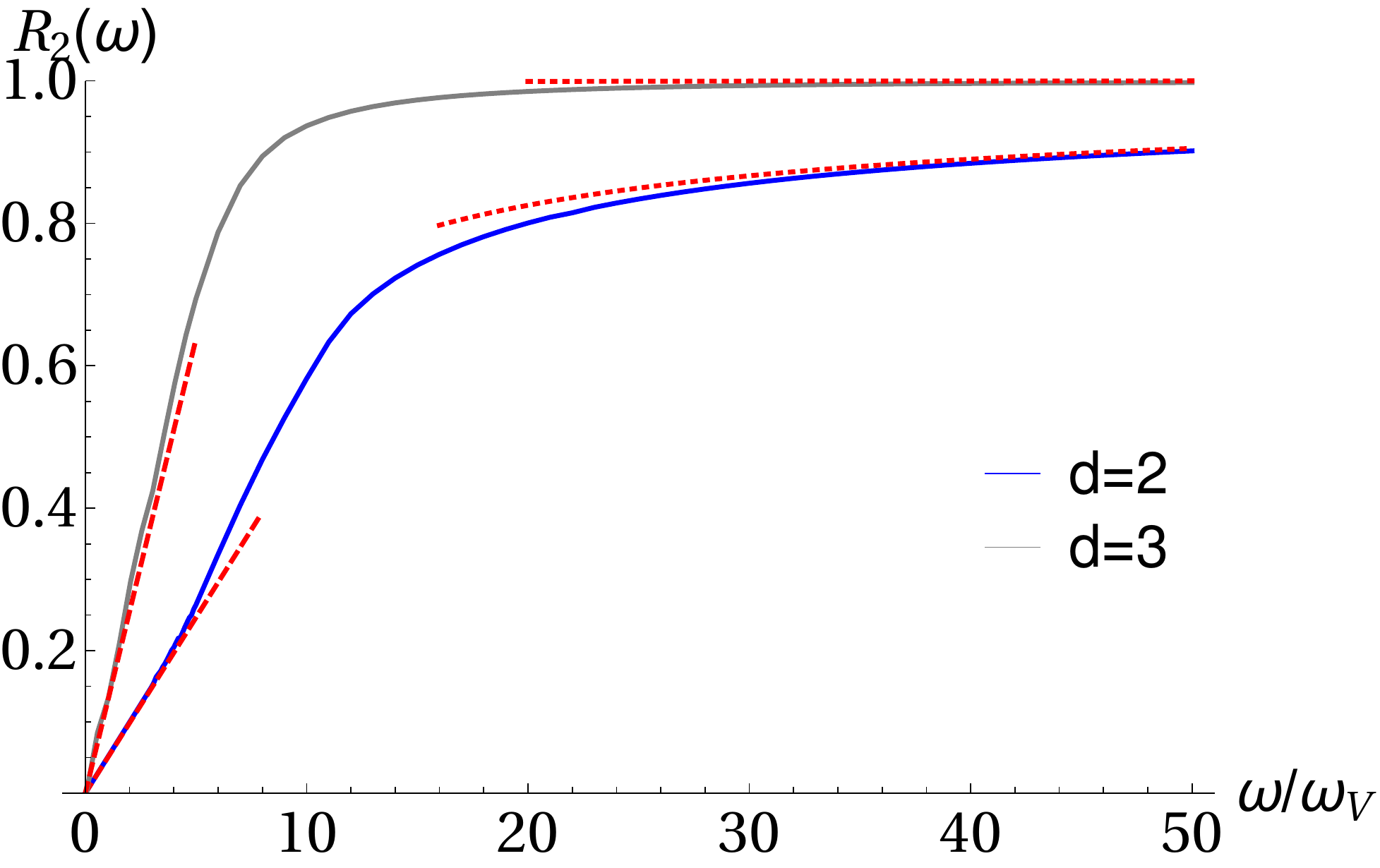} 
\caption{\label{fig:R2(y)}
(Colour online)
The correlation function $R_{2}(\omega)$ of the dipole energy levels in an electronic system
as a function of frequency $\omega$ [in units $\omega_V$], obtained from a numerical integration of Eq.~(\ref{eq:R2_interm1}).
Red dashed lines indicate the small- and large- frequency asymptotic behaviour  
given by Eqs. (\ref{eq:d_2_behavior}) and (\ref{eq:R2_large_omega_d=3_strong_disorder}) are shown in red dashed lines. 
} 
\end{figure}

The whole frequency dependency of the correlation function $R_2(\omega)$ for dimensions
$d=2$ and $d=3$, obtained
from numerical integration of Eq.~(\ref{eq:R2_interm1}), is shown in Fig.~\ref{fig:R2(y)}.
The dashed lines in Fig.~(\ref{fig:R2(y)}) show the low-frequency behaviour described by the
linear dependence (\ref{R2omega}); the dotted lines
show the high-frequency asymptotics described by 
Eqs.~(\ref{eq:d_2_behavior}) and (\ref{eq:R2_large_omega_d=3_strong_disorder}).

\section{Arbitrary disorder strength in a 3D electronic system}
\label{Sec:Weaker_Disorder}

In Sec.~\ref{Sec:StrongDis} we considered energy-level correlations in a system, which is either sufficiently
strongly disordered or sufficiently small, and only sparse resonances between pairs of dipole
sites are essential for correlations of energy levels, while higher-order resonances may be neglected.
As discussed in Sec.~\ref{Sec:StrongDis} and as first pointed out in
Refs.~\onlinecite{Levitov_1990,Levitov:AnnReview,Levitov2},
excitations with the hopping amplitude $\propto1/r^3$ have infinitely many resonances in 3D systems
at arbitrarily strong disorder, unlike systems in lower dimensions $d<3$. Therefore, 
sufficiently large 3D electronic systems may host rather complicated dipole states, which involve resonances on multiple sites.

The correlation functions may then be still found by replacing the dipole states by effective
hybridised states on the scale of the volume $V$ of interest and investigating the hopping matrix elements for
such states. The energies of the effective hybridised states may be assumed to have a uniform 
probability distribution due to the uniform energy distributions of dipoles on constituent sites and their
independence of the hopping elements.
Assuming that the hybridised states have dipole moments, one may expect the results of Sec.~\ref{Sec:StrongDis}
for the frequency dependency of the correlator $R_2(\omega)$ 
to carry over directly to the case of weaker disorder, which allows for complicated hybridised states.
Namely, one may expect that at low frequencies $R_2(\omega)\propto \omega$, while for $\omega\rightarrow\infty$
$R_2(\omega)=1-\frac{\text{const}}{\omega^2}$.
We emphasise, however, that the dependencies on the system size (volume $V$) may be different from those
found in Sec.~\ref{Sec:StrongDis} because the effective dipole moments of hybridised states of multiple sites
in general depend on the volume $V$.

Three decades ago, a renormalisation procedure was developed in Ref.~\onlinecite{Levitov_1990}
for constructing hybridised dipole-like states in 3D by repeatedly hybridising pairs of dipole-like states
with closest energies within a given distance while increasing the system size or the interaction
radius.
A more rigorous recent study\cite{AleinerEfetov:dipoles} for a similar 2D problem with $\propto 1/r^2$ dipole
hopping established the existence of fixed points in a system on the orthogonal symmetry class (which is also the focus
of this paper)
with critical wavefunctions of the dipole states.

In what immediately follows we construct a renormalisation procedure at sufficiently strong disorder, similar to that
of Ref.~\onlinecite{Levitov_1990}, to explore qualitatively the correlations of complex multi-site excitations
with effective renormalised dipole moments and recombination matrix elements $J_{\br\bd}
\equiv \bra{0}\mathcal{\hat H}_0\ket{\br\bd}$ (the matrix elements
of the Hamiltonian between the excitation state and the ground state of a non-interacting system).

When two dipole states $(\br\bd)$ and $(\br^\prime\bd^\prime)$ are hybridised,
they are being replaced by two other states with the annihilation operators
$ \hb_{\br_+\bd_+}$ and  $\hb_{\br_-\bd_-}$
\begin{align}
	\left(
	\begin{array}{c}
		\hb_{\br_+\bd_+} \\
		\hb_{\br_-\bd_-}
	\end{array}
	\right)
	=
	\left(\begin{array}{cc}
	\cos \theta &\sin \theta\\
	-\sin \theta &\cos \theta 
	\end{array}\right)
	\left(
	\begin{array}{c}
		\hb_{\br\bd} \\
		\hb_{\br^\prime\bd^\prime}
	\end{array}
	\right),
\end{align}
where $\cot(2\theta)\equiv\tau=(E_{\br\bd}-E_{\br'\bd'})/2T_{\br\bd,\br'\bd'}$.
The recombination matrix elements of the hybridised states and the elements of hopping to remote sites 
$\tilde\br\tilde{\bd}$ with $|\tilde{\br}-\br|,|\tilde{\br}-\br'|\gg |\br-\br'|$ are given by 
\begin{align}
&(J_{\br_+\bd_+},J_{\br_-\bd_-})^T=U(J_{\br\bd},J_{\br'\bd'})^T, \\
&(T_{\tilde{\br}\tilde{\bd},\br_+\bd_+},T_{\tilde{\br}\tilde{\bd},\br_-\bd_-})^T=U(T_{\tilde{\br}\tilde{\bd},\br\bd},T_{\tilde{\br}\tilde{\bd},\br'\bd'})^T. 
\end{align}

Because the hopping of dipole excitations depends on the product $J_{\br\bd}\equiv\bp_{\br\bd}\bd$
of its recombination element $J_{\br\bd}$ and the dipole moment $\bd$, it is convenient to introduce
a new variable 
\begin{align}
	\bp_{\br\bd}\equiv J_{\br\bd}\bd
\end{align}
and describe the evolution of its distribution
function $F(\bp_{\br\bd})$ when repeatedly hybridising dipole states.
Assuming that the initial distributions of the dipole moments $\bd$ and the 
recombination elements are isotropic, the distribution $F(\bp_{\br\bd})$ also remains isotropic
under renormalisation and depends only on the absolute value of $\bp_{\br\bd}$. 

The renormalisation procedure involves repeated hybridisation of pairs of dipole states
with close energies. When increasing the system size $L$ (or the interaction radius), new states
are formed out of previously hybridised states.
In the spirit of Ref.~\onlinecite{Levitov_1990}, we
neglect resonances of three or more sites, which is justified in the limit of sufficiently 
strong disorder under consideration, with
$ \nu_0 \langle\bd^2\rangle J^2 E_{\br\bd}^{-2} n \ll 1$, where $n=N/V$ is the density of the dipoles.
Introducing variable $\ell=\log L$, where $L$ is the system size (or the
interaction cutoff radius), we arrive at the RG flow equation
for the distribution function $F(\bp_{\br\bd})$:
\begin{align}
&\frac{\partial F \left(\bp_{\br\bd}\right)}{\partial\ell}= \frac{n\nu_0}{E^2}\int d\bn \int d\bp_{\br_1\bd_1} d\bp_{\br_2\bd_2} F\left(\bp_{\br_1\bd_1}\right)F\left(\bp_{\br_2\bd_2}\right)\nonumber\\
&\int d\tau\left|Q\left(\bp_{\br_1\bd_1},\bp_{\br_2\bd_2},\bn\right)\right|
\left[\delta\left(\bp-\bp_{+}\right)+\delta\left(\bp-\bp_{-}\right)-\right. \nonumber\\
&\hspace{1.5 in}\left.\delta\left(\bp-\bp_{\br_1\bd_1}\right)-\delta\left(\bp-\bp_{\br_2\bd_2}\right)\right].
\label{Frenorm}
\end{align}
Similarly, when increasing the size of the system (or the interaction cutoff radius), the 
energy-level correlation function $R_2(\omega)$ gets renormalised according to the equation
(see Appendix for details)
\begin{widetext}
\begin{align}
 \frac{\partial R_2(\omega, \ell)}{\partial \ell}=\frac{1}{V}\int d\bn  \int d\bp_{\br_1\bd_1} \ d\bp_{\br_2\bd_2}F\left(\bp_{\br_1\bd_1}\right)F \left(\bp_{\br_2\bd_2}\right)\int d\tau 
 \left|Q\left(\bp_{\br_1\bd_1},\bp_{\br_2\bd_2},\bn\right)\right|/E^2\nonumber\\
 \hspace{2. in}\left[\delta\left(\omega-2\left|T_{{\br_1\bd_1,\br_2\bd_2}}\right|\sqrt{\tau^{2}+1}\right)-2\delta\left(\omega-2T_{{\br_1\bd_1,\br_2\bd_2}}{\tau}\right)\right].
 \label{eq:R_2_differential}
\end{align}
\end{widetext}
The solutions of the RG Eqs.~(\ref{Frenorm}) and (\ref{eq:R_2_differential})
describe the distributions of the parameters of dipole states and energy-level correlation functions
in a 3D system of arbitrary size.

{\it Level correlations at intermediate lengths.} Because the hybridised states are very close in structure
to large-size dipoles, it is possible to apply immediately the results of Sec.~\ref{Sec:StrongDis}
to their energy-level corrections, which gives
\begin{align}
	R_2(\omega)=
	\left\{
	\begin{array}{cc}
		\tilde{C}_3(V) \omega,& \omega\rightarrow 0 \\
		1-{\tilde{C}_2(V)}/{\omega^2}, & \omega\rightarrow\infty.
	\end{array}
	\right.
	\label{R2Large3D}
\end{align}
We note, however, that the dependencies of the coefficients $\tilde{C}_3(V)$
and $\tilde{C}_2(V)$ on the system of volume $V$ are in general different from the dependencies 
$\tilde{C}_3(V)\propto V$ and {$\tilde{C}_2(V)\propto 1/V^2$} in the case of strong disorder [cf. Eqs.~(\ref{R2omega})
and (\ref{eq:R2_large_omega_d=3_strong_disorder})],
because the parameters of the hybridised composite dipoles depend on the system size. 
These dependencies will depend on the details of the initial dipole distributions.

{\it Fixed point.} The procedure of the hybridisation described above leaves invariant
the quantities $J_{\br_1\bd_1}^2+J_{\br_2\bd_2}^2$ and $\bp_{\br_1\bd_1}^2+\bp_{\br_2\bd_2}^2$
for pairs of states. As a result, the quantity $\bp_{\br\bd}$ and the dipole states remains bounded for
the typical renormalised composite dipole states. It is natural to assume then that the distributions
of the parameters $p_{\br\bd}$ of the renormalised dipole states approach a fixed point 
$F^*(\bp_{\br\bd})$. Such type of a fixed point has been obtain, under certain approximations,
in Ref.~\onlinecite{Levitov_1990} for a similar model for dipoles with the power-law hopping $\propto 1/r^3$
in 3D. Assuming a similar fixed point exists here, the typical value of the parameter $\bp_{\br\bd}$
saturates to a constant value in sufficiently large sample, and the results of Sec.~(\ref{Sec:StrongDis})
for both the energy and size dependencies of the energy-level correlator $R_2(\omega)$ may be carried 
over directly. Thus, the energy-level statistics in very large 3D systems
are given by Eq.~(\ref{R2Large3D}) with $\tilde{C}_3(V)\propto V$ and {$\tilde{C}_2(V)\propto 1/V^2$}.

\section{Generic power-law hopping}

\label{Sec:GenericPowerLaw}



Excitations with power-law hopping are often simulated by means of ultracold particles in magnetic or
optical traps. Recently, excitations with power-law hopping $\propto 1/r^\alpha$ with tunable $\alpha=0\ldots3$
have been 
realised in 1D\cite{Monroe:longrange,Islam:longrange,Blatt:chain1,Blatt:chain2} and
2D\cite{Bollinger:longrange} arrays of trapped ultracold ions. 
Excitations with power-law hopping also exist in systems of Rydberg atoms\cite{Rydberg_Review_2010}
($\alpha=6$ or $\alpha=3$) and polar molecules\cite{dipolar_experiment_2013,dipolar_review_2009} ($\alpha=3$).

In what follows, we compute energy-level statistics in a strongly disordered system with a generic
power-law hopping described by the Hamiltonian
\begin{eqnarray}
&&\mathcal{H}  =
\sum_{\br,\lambda}E_{\br\lambda^\prime}
\hb_{\br\lambda}^{\dagger}\hb_{\br\lambda^\prime}
-\sum_{\br,\br^\prime,\lambda,\lambda^\prime}
T_{\br\lambda,\br^\prime\lambda^\prime}\hb_{\br\lambda}^\dagger\hb_{\br^\prime\lambda^\prime}
\end{eqnarray}
where the operators $\hb_{\br\lambda}$ and $\hb_{\br\lambda}^\dagger$ annihilate and create
excitations at location $\br$; $\lambda$ labels discrete degrees of freedom of excitations
at a given location, e.g. the spatial orientation
of the excitations; and we have also introduced  
the hopping element
\begin{align}
	T_{\br\lambda,\br^\prime\lambda^\prime}=\frac{2Q_{\alpha}(\lambda,\lambda^\prime)}{|\br-\br^\prime|^\alpha}.
\end{align}
with the kernel $Q_{\alpha}(\lambda,\lambda^\prime)$ which is independent of the distance but depends
on the excitation states $\lambda$.


{\it Correlations at strong disorder.}
In what follows, we compute the energy-level correlation function in the case of very strong disorder,
when correlations come from rare resonances on pairs of sites. Following the same steps as when deriving
Eq.~(\ref{eq:R2_interm1}), we arrive at
\begin{align}
R_{2}\left(\omega\right)=&\frac{1}{V}\int d\lambda\ \int d\lambda^{\prime}\ P\left(\lambda\right)P\left(\lambda^{\prime}\right)\nonumber \\ 
& \int_{\tilde{r}>\tilde{r}_{\omega}}d\tilde{\boldsymbol{r}}\left[1-\left(\frac{2Q_{\alpha}\left(\lambda,\lambda^{\prime},\tilde{\boldsymbol{r}}/\tilde{r}\right)}{\omega\tilde{r}^{\alpha}}\right)^{2}\right]^{-\frac{1}{2}}\label{eq:R2_higher_exc}
\end{align}
where $\tilde{r}_{\omega}=\left({2Q_{\alpha}}/{\omega}\right)^{\frac{1}{\alpha}}$
and $P(\lambda)$ is the probability distribution of the excitation states $\lambda$.

The typical splitting between neighbouring energy levels is given by the 
characteristic matrix element of quasiparticle hopping 	
\begin{align}
	\omega_V=2 V^{-\frac{\alpha}{d}}\left<|Q(\lambda,\lambda^\prime)|\right>_{\lambda,\lambda^\prime}
\end{align}
where $\langle\cdots\rangle_\lambda=\int d\lambda P(\lambda)\cdots$.
For sufficiently smooth probability distributions of the function $Q(\lambda,\lambda^\prime)$
and its moments $Q(\lambda,\lambda^\prime)^\beta$, the coefficients
\begin{align}
	\mathcal{A}_\beta=
	\frac{2^\beta\left<|Q(\lambda,\lambda^\prime)|^\beta\right>_{\lambda,\lambda^\prime}}{V^\frac{\alpha\beta}{d}\omega_V^\beta}
	\label{Abeta}
\end{align}
are of order unity, and $\omega_V$ is the only energy scale in the problem.
In the limit $\omega\gg\omega_V$ the correlation function is given by
\begin{align}
R_{2}\left(\omega\gg\omega_V\right)\approx 1-\frac{2\pi^{\frac{d+1}{2}}\Gamma\left(1-\frac{d}{2\alpha}\right)\mathcal{A}_\frac{d}{\alpha}}{\Gamma\left(\frac{d}{2}\right)\Gamma\left(\frac{1}{2}-\frac{d}{2\alpha}\right)d}\left(\frac{\omega_{V}}{\omega}\right)^{\frac{d}{\alpha}}.
\label{R2AlphaPre}
\end{align}

Due to the divergence of the Gamma function $\Gamma\left(\frac{1}{2}-\frac{d}{2\alpha}\right)\sim\frac{2\alpha}{\alpha-d}$ 
when the dimension $d$ approaches $\alpha$, the coefficient in the last term in Eq.~(\ref{R2AlphaPre})
vanishes for $d=\alpha$.
Therefore, the energy-level correlations in the dimensions $d=\alpha$ at $\omega=\omega_V$
are described by the next-leading term in $1/\omega$:
\begin{align}
	R_{2}^{d=\alpha}\left(\omega\gg\omega_{V}\right)\approx1-A_{2}
	\left(\frac{\omega_V}{\omega}\right)^2,
	\label{R2AlphaD}
\end{align}
where $A_2$ is a coefficient of order unity which depends on the shape of the system.
Equations (\ref{R2AlphaPre}) and (\ref{R2AlphaD}) may be combined into one interpolation formula 
\begin{align}
R_{2}\left(\omega\right)\approx
1-\left(\alpha-d\right)A_{1}\left(\frac{\omega_V}{\omega}\right)^\frac{d}{\alpha}
-A_{2}\left(\frac{\omega_V}{\omega}\right)^2.
\label{Interpolation3}
\end{align}

{\it Small-frequency limit.} In the limit $\omega\ll\omega_V$ we obtain, similarly to Eq.~(\ref{R2omega}),
\begin{equation}
R_{2}\left(\omega\right)=A_{3}\frac{\omega}{\omega_{V}},
\label{R2AlphaSmallOmega}
\end{equation}
where the dimensionless coefficient $A_{3}=\frac{\pi\left\langle\left|Q(\lambda,\lambda^{\prime})\right|\right\rangle_{\lambda,\lambda^{\prime}}}{V^{1+\frac{\alpha}{d}}}\int d\tilde{\boldsymbol{r}}\tilde{r}^{\alpha}\int d\lambda d\lambda^{\prime}P\left(\lambda\right)P\left(\lambda^{\prime}\right)\delta(Q_{\alpha})$ depends on the shape of the system.

{\it Behaviour in high and low dimensions.}
Equations (\ref{Interpolation3}) and (\ref{R2AlphaSmallOmega}) describe energy-level correlations accurately in
low dimensions $d<\alpha$ or in sufficiently small systems in higher dimensions $d\geq\alpha$.
In that case, the correlations come from rare resonances between excitation states on 
pairs of sites. In higher dimensions, $d\geq\alpha$, the number of resonances is infinite in the limit
of an infinite system, which may lead to a strong renormalisation
of the excitation states.
 Based on the arguments similar to those of Sec.~\ref{Sec:Weaker_Disorder}, we expect that
in higher-dimensional systems the frequency dependency of the correlation function
is still given by Eqs.~(\ref{Interpolation3}) and (\ref{R2AlphaSmallOmega}), however,
the volume dependence is in general different.

\section{Conclusion and outlook}

\label{Sec:conclusion}

Motivated by neutral excitations in disordered electronic systems and trapped ultracold particles
with power-law interactions, we have computed the energy-level correlations functions for particles
with the power-law hopping $\propto r^\alpha$. Our main results for the correlation functions for systems
in various dimensions and various energy intervals are summarised by Eqs.~(\ref{R2StrongGen}), 
(\ref{eq:R2_large_omega_d=3_strong_disorder}), (\ref{R2omega}), (\ref{R2AlphaPre}), (\ref{R2AlphaD})
and (\ref{R2AlphaSmallOmega}).
In a disordered electronic systems the correlation function may be observed 
as a correlator of ac conductances
$R_2 (\omega) \propto \left<G(\Omega)G(\Omega+\omega)\right>_\Omega$,
where $\langle\ldots\rangle_\Omega$ is the averaging with respect to frequency $\Omega$
in a sufficiently large energy window.

At small energy differences, the energy-level correlations displays Wigner-Dyson statistics,
which hints at the possibility of chaotic dynamics of the excitations involved. This chaotic
behaviour could be identified, for example, via out-of-time-order 
correlators\cite{LarkinOvchnnikov} of, e.g., 
local voltages or charges in a system with excitations which allow for power-law hopping.
We do not present such analysis here and leave them for future studies.

Another question, which deserves a separate study, is the relation between the disorder strength 
and the level statistics in systems with the power-law hopping $\propto1/r^\alpha$ for sufficiently small $\alpha$.
Indeed, those system support excitations with the dispersion $\varepsilon_\bk\propto k^{\alpha-d}$.
In dimensions $d>\frac{3\alpha}{2}$ they display a plenty of unconventional disorder-driven phenomena,
such as disorder-driven transitions or sharp crossovers in non-Anderson universality classes,
unconventional Lifshitz tails, etc., (see Ref.~\onlinecite{Syzranov:review} for a review) and possibly
transitions in the energy-level statics. While we have obtained the strong-disorder asymptotics of the
respective level statistics in this paper, we leave further studies of the possibility of such transitions
for future work.

\section{Acknowledgements}
The authors would like to thank Alexey Gorshkov for useful discussions.
PT and SVS have been financially supported by NSF QIS, AFOSR, NSF PFC at JQI, ARO MURI, ARO and ARL CDQI. PT is supported by the NRC postdoctoral fellowship.
SVS also acknowledges start-up funds at the University of California at Santa Cruz.
VLQ acknowledges financial support from the National High Magnetic Field Laboratory through NSF Grant No. 
DMR-1157490 and the State of Florida, and thanks the Aspen Center for Physics, supported by the
National Science Foundation through grant PHY-1607611, for hospitality. 


\bibliography{references} 


\onecolumngrid


\appendix

\section{Change to the level correlation function when adding hopping between two sites}
\label{Sec:R2modificationStep}

In this section we derive the modification of the correlation function $R_2(\omega)$,
defined by Eq. \ref{R2definiiton},
when two dipole states hybridise.
We consider a system of $N$ sites, where two sites $(\br_1,\bd_1)$ and $(\br_2,\bd_2)$
are initially isolated
from each other and from the rest of the system and compute the change of $R_2(\omega)$ when adding
hopping $T_{\br_1\bd_1,\br_2\bd_2}$ between the two sites.
The model under consideration applies to the case of very strong disorder, when resonant pairs of sites are rare, as well
as to dipole hybridisation during one step of the strong-disorder RG, when composite dipole states may be considered,
and the hybridisation between two dipole states being merged and the rest of the system may be neglected.

The modified density of dipole states after adding the hopping
\begin{equation}
\tilde{\nu}(E)=\left(\delta(E-E_+)+\delta(E-E_-)+\sum_{\lambda}\delta(E-E_{\lambda})\right)
\label{NuModified}
\end{equation}
consists of the contribution of the hybridised-states' energies $E_+$ and $E_-$ and that of the rest of the system
[the sum in Eq.~(\ref{NuModified})]. Because the hopping is small, and the hopping and the on-site dipole
energies fluctuate independently, the average density of states is unaltered by the hybridisation
 $\langle\tilde{\nu}\rangle_{\rm dis}=\langle \nu\rangle_{\rm dis}=N\nu_0$.
The change of to the disorder-averaged correlation function $R_2(\omega)$, defined by Eq.~(\ref{R2definiiton}),
is given by
\begin{align}
\delta R_2(\omega)&=&\frac{1}{N^2\nu_0^2} \int d J_{{\br_1\bd_1}} P(J_{{\br_1\bd_1}})\int d J_{\br_2\bd_2}P(J_{\br_2\bd_2}) \int d\bd_1\  d\bd_2\ f\left(\bd_1\right)f\left(\bd_2\right)\int \nu_0 dE_{{\br_1\bd_1}}\nu_0 dE_{{\br_2\bd_2}}\nonumber\\
&&\hspace{0in} 
\left<\tilde{\nu}\left(E+\frac{\omega}{2}\right)\tilde{\nu}\left(E-\frac{\omega}{2}\right)-\nu\left(E+\frac{\omega}{2}\right)\nu\left(E-\frac{\omega}{2}\right)\right>_\lambda,
\label{DensityRProducts}
\end{align}
where $\langle\ldots\rangle_\lambda$ is the averaging with respect to the disorder in the rest of the system,
independent of the parameters on sites $(\br_1,\bd_1)$ and $(\br_2,\bd_2)$.

We note that, according to Eq.~(\ref{NuModified}), products of the modified densities of states
$\tilde{\nu}\left(E+\frac{\omega}{2}\right)\tilde{\nu}\left(E-\frac{\omega}{2}\right)$
in Eq.~(\ref{DensityRProducts})
contain three types of terms involving products of $\delta$-functions,
(i) $\delta(E-E_\pm+\omega/2)\delta(E-E_\mp-\omega/2)$, (ii) $\delta(E-E_\pm+\omega/2)\delta(E-E_{\lambda}-\omega/2)$
and (iii)
$\delta(E-E_{\lambda}+\omega/2)\delta(E-E_{\lambda'}-\omega/2)$.
Terms (iii) are cancelled by equivalent 
contributions from $\nu\left(E+\frac{\omega}{2}\right)\nu\left(E-\frac{\omega}{2}\right)$.
Contributions of type (ii) also vanish, due to the identity
\begin{align}
\int \,dE_{{\br_1\bd_1}}\, dE_{{\br_2\bd_2}}
\left[\delta(E-E_+)+\delta(E-E_-)-\delta(E-E_{{\br_1\bd_1}})-\delta(E-E_{{\br_2\bd_2}})\right]=0.
\label{eq:constantDOSidentity}
\end{align}
Equation (\ref{DensityRProducts}) may therefore be simplified to include only the averaging with respect to 
the dipole parameters on sites $(\br_1,\bd_1)$ and $(\br_2,\bd_2)$:
\begin{eqnarray}
\delta R_2(\omega)&=&\frac{1}{N^2} \int d J_{{\br_1\bd_1}} P(J_{{\br_1\bd_1}})\int d J_{\br_2\bd_2}P(J_{\br_2\bd_2})\int d\bd_1\  d\bd_2\ f\left(\bd_1\right)f\left(\bd_2\right)\int dE_{{\br_1\bd_1}}\int dE_{{\br_2\bd_2}}\nonumber\\
&&\left\{ \left[\delta\left(E+\frac{\omega}{2}-E_{+}\right)+\delta\left(E+\frac{\omega}{2}-E_{-}\right)\right]
\left[\delta\left(E-\frac{\omega}{2}-E_{+}\right)+\delta\left(E-\frac{\omega}{2}-E_{-}\right)\right]- \right.\nonumber\\  
&&\left.\left[\delta\left(E+\frac{\omega}{2}-E_{{\br_1\bd_1}}\right)+\delta\left(E+\frac{\omega}{2}-E_{{\br_2\bd_2}}\right)\right]\left[\delta\left(E-\frac{\omega}{2}-E_{{\br_1\bd_1}}\right)+\delta\left(E-\frac{\omega}{2}-E_{{\br_2\bd_2}}\right)\right]\right\}.
\end{eqnarray} Changing variables to $\tau=\frac{1}{2T_{\boldsymbol{r,d},\boldsymbol{r'd'}}}\left(E_{\boldsymbol{rd}}-E_{\boldsymbol{r'd'}}\right)$ and  integrating out $\frac{1}{2}(E_{\boldsymbol{rd}}+E_{\boldsymbol{r'd'}})$ gives
  \begin{align}
\delta R_2(\omega) &=& \frac{2}{N^2} \int d J_{{\br_1\bd_1}} P(J_{{\br_1\bd_1}})\int d J_{\br_2\bd_2}P(J_{\br_2\bd_2}) \int d\bd\ d\bd_2f\left(\bd_1\right)f \left(\bd_2\right)\times\nonumber\\
&&\times\int d\tau\, |T_{{\br_1\bd_1},{\br_2\bd_2}}| \left[\delta\left(\omega-2\left|T_{{\br_1\bd_1,\br_2\bd_2}}\right|\sqrt{\tau^{2}+1}\right)-2\delta\left(\omega - 2\tau T_{{\br_1\bd_1},{\br_2\bd_2}}\right)\right].
\label{eq:dR_2_pert}
 \end{align}
 So far, we have obtained the expression for the correction to the level correlation, following the hybridization of a single pair of dipoles. It is straightforward to obtain the expression for the level correlation given in Eq. \ref{eq:R_2_Main} which takes into account all resonances in a given volume $V$ by integrating over $\frac{1}{2}\frac{N^2}{V^2}d\br_1d\br_2$ and then integrating out the variable $\tau$.
 

\subsection*{Application to the strong-disorder renormalisation procedure}

The modification of the correlation function $R_2(\omega)$ when hybridising dipole states may be considered
as a step of an RG procedure, discussed in Sec.~\ref{Sec:Weaker_Disorder} and applied in
Refs.~\onlinecite{Levitov_1990,Levitov:AnnReview,Levitov2} (see also Ref.~\onlinecite{MirlinEvers:R2}).
During this procedure, pairs of resonant dipole states 
are being repeatedly hybridised while increasing the hopping distance 
$r=|\boldsymbol{r_1}-\boldsymbol{r_2}|=e^\ell$ 
or the system size.

As discussed in Section.~\ref{Sec:Weaker_Disorder}, the hopping of dipoles depends only on the product 
$\bp_{\br\bd}=J_{\br\bd}\bd$, which is why it is convenient to introduce the distribution function $F(\bp_{\br\bd})$
of variable $\bp_{\br\bd}$, which flows under the renormalisation procedure. We can now obtain the modification to $R_2(\omega)$ as a result of hybridizing dipoles in the volume element. 
The number of dipole pairs separated by vectors $\br$ in an infinitesimal element of space, confined by the radii
$r$ and $r+dr$ and the spatial angle $d\Omega$, is given by 
 $\frac{1}{2} r^2dr\int d\Omega\int \frac{N}{V}d\br_1 \frac{N}{V} d\br_2\delta\left[\br-\left(\boldsymbol{r_1}-\boldsymbol{r_2}\right)\right]$.
Utilizing the expression in Eq.~(\ref{eq:dR_2_pert}) for hybridization of  a single pair of dipoles and multiplying by the number of dipoles in the volume element, we obtain the modification of the energy-level correlation function
\begin{align}
\delta R_2(\omega,\ell) =& r^2dr\int d\Omega\int \frac{1}{2}\frac{N}{V}d\br_1 \frac{N}{V} d\br_2\delta\left[\br-\left(\boldsymbol{r_1}-\boldsymbol{r_2}\right)\right]\cdot \frac{1}{N^2}\ \int d\bp_{\br_1\bd_1}\ d\bp_{\br_2\bd_2} F\left(\bp_{\br_1\bd_1}\right)F \left(\bp_{\br_2\bd_2}\right)
\nonumber\\
& 
2\int d\tau\, |T_{{\br_1\bd_1},{\br_2\bd_2}}| \left[\delta\left(\omega-2\left|T_{{\br_1\bd_1,\br_2\bd_2}}\right|\sqrt{\tau^{2}+1}\right)-2\delta\left(\omega - 2\tau T_{{\br_1\bd_1},{\br_2\bd_2}}\right)\right]
\label{deltaR2pre}
 \end{align}
where the factor of $\frac{1}{2}$ in the right-hand side prevents
double counting of dipoles.

To make further progress, we note that the contribution to Eq.~(\ref{deltaR2pre}), which comes from 
the second $\delta-$function, may be simplified as
  \begin{equation}
 \int \frac{d\br}{2}\int \frac{d\br_1\   d\br_2}{V^2}\delta\left[\br-\left(\boldsymbol{r_1}-\boldsymbol{r_2}\right)\right] \int d\bp_{\br_1\bd_1} \ d\bp_{\br_2\bd_2}F\left(\bp_{\br_1\bd_1}\right)F \left(\bp_{\br_2\bd_2}\right) 
 \int d\tau\cdot 2|T_{{\br_1\bd_1},{\br_2\bd_2}}| \left[-2\delta\left(\omega - 2\tau T_{{\br_1\bd_1},{\br_2\bd_2}}\right)\right]=-1.
 \end{equation}
 The correlation function $R_2(\omega)$ is obtained by integrating Eq.~(\ref{deltaR2pre})
with respect to $\ell$ from $\ell=0$ to $\ell=\log L$, where $L$ is the size of the system or the interaction
cutoff radius.
Performing also integration with respect to $(\boldsymbol{r_1}-\boldsymbol{r_2})$ and $\frac{1}{2}(\boldsymbol{r_1}+\boldsymbol{r_2})$ over the volume $V$, we arrive at
 \begin{equation}
 \int_{\ell=0}^{\ell=\ln L}\frac{\partial R_2(\omega, \ell)}{\partial\ell}d\ell
 =-1+\frac{1}{V}\int d^3r  \int d\bp_{\br_1\bd_1} \ d\bp_{\br_2\bd_2}F\left(\bp_{\br_1\bd_1}\right)F \left(\bp_{\br_2\bd_2}\right)\int d\tau\, |T_{{\br_1\bd_1},{\br_2\bd_2}}| \delta\left(\omega-2\left|T_{{\br_1\bd_1,\br_2\bd_2}}\right|\sqrt{\tau^{2}+1}\right).
 \label{R2integral}
 \end{equation}
Equation (\ref{R2integral}) together with the initial condition $R_2(\omega,\ell=0)=1$ gives
 \begin{equation}
   R_2(\omega, \ln L)=\frac{1}{V}\int_0^{\ln L} d\ell\int d\Omega  \int d\bp_{\br_1\bd_1}\ d\bp_{\br_2\bd_2} F\left(\bp_{\br_1\bd_1}\right)F \left(\bp_{\br_2\bd_2}\right)\int d\tau 
   \frac{\left|Q\left(\bp_{\br_1\bd_1},\bp_{\br_2\bd_2}\right)\right|}{E^2}
   \delta\left(\omega-2\left|T_{\br_1\bd_1,\br_2\bd_2}\right|\sqrt{\tau^{2}+1}\right).
   \label{R2almostMain}
 \end{equation}
Equation (\ref{R2almostMain}) may also be rewritten in the form of the RG flow equation 
\begin{eqnarray}
 \frac{\partial R_2(\omega, \ell)}{\partial \ell}&=&\frac{1}{V}\int d\Omega  \int d\bp_{\br_1\bd_1} \ d\bp_{\br_2\bd_2}F\left(\bp_{\br_1\bd_1}\right)F \left(\bp_{\br_2\bd_2}\right)\int d\tau \frac{\left|Q\left(\bp_{\br_1\bd_1},\bp_{\br_2\bd_2},\Omega\right)\right|}{E^2}\nonumber\\
&&\hspace{2. in}\left[\delta\left(\omega-2\left|T_{{\br_1\bd_1,\br_2\bd_2}}\right|\sqrt{\tau^{2}+1}\right)-2\delta\left(\omega-2T_{{\br_1\bd_1,\br_2\bd_2}}{\tau}\right)\right].
\end{eqnarray}

\section{Change to the distribution function of dipoles when hybridising two sites}

In this section, we derive the RG flow equation for the distribution function
$F(\bp_{\br\bd})$ of the dipole parameter $\bp_{\br\bd}=J_{\br\bd}\bd$, the product of the dipole moment $\bd$
and the recombination matrix element $J_{\br\bd}$, discussed in Sec.~\ref{Sec:Weaker_Disorder}.
When two dipoles with parameters $\bp_{\br_1\bd_1}$ and $\bp_{\br_2\bd_2}$ are hybridised,
they get replaced by two other dipole states with parameters $\bp_+$ and $\bp_{-}$,
and the distribution function gets modified according to 
\begin{align}
\delta F \left(\bp_{\br\bd}\right)=& \int d\bp_{\br_1\bd_1} d\bp_{\br_2\bd_2} F\left(\bp_{\br_1\bd_1}\right)F\left(\bp_{\br_2\bd_2}\right)\int \nu_0 d E_{\br_1\bd_1}\nu_0 d E_{\br_2\bd_2}\nonumber \\ 
&\hspace{2.5 in}\left[\delta\left(\bp-\bp_{+}\right)+\delta\left(\bp-\bp_{-}\right)-
\delta\left(\bp-\bp_{\br_1\bd_1}\right)-\delta\left(\bp-\bp_{\br_2\bd_2}\right)\right]
\end{align}
Changing variables to $\tau=(E_{\br_1\bd_1}-E_{\br_2\bd_2})/\left(2T_{\br_1\bd_1,\br_2\bd_2}\right)$, and considering the effects of all resonances in a spherical shell of radius $r\rightarrow r+dr$, with $r=\left|\br_1-\br_2\right|$, gives
\begin{align}
\delta F \left(\bp_{\br\bd}\right)=& \frac{N}{V}r^2dr\int d\Omega \int d\bp_{\br_1\bd_1} d\bp_{\br_2\bd_2} F\left(\bp_{\br_1\bd_1}\right)F\left(\bp_{\br_2\bd_2}\right)\int \nu_0 \left|T_{\br_1\bd_1,\br_2\bd_2}\right|d\tau \nonumber \\ 
&\hspace{2.5 in}\left[\delta\left(\bp-\bp_{+}\right)+\delta\left(\bp-\bp_{-}\right)-
\delta\left(\bp-\bp_{\br_1\bd_1}\right)-\delta\left(\bp-\bp_{\br_2\bd_2}\right)\right].
\end{align}
Introducing the RG parameter $\ell=\log r$
and using that 
 $T_{\br_1\bd_1,\br_2\bd_2}=Q(\bp_{\br_1\bd_1},\bp_{\br_2\bd_2})/\left(E^2r^3\right)$, 
 we obtain the RG flow equation for the distribution function $F\left(\bp_{\br\bd}\right)$:
\begin{align}
&\frac{\partial F \left(\bp_{\br\bd}\right)}{\partial\ell}= \frac{n\nu_0}{E^2}\int d\Omega \int d\bp_{\br_1\bd_1} d\bp_{\br_2\bd_2} F\left(\bp_{\br_1\bd_1}\right)F\left(\bp_{\br_2\bd_2}\right)\int d\tau\left|Q\left(\bp_{\br_1\bd_1},\bp_{\br_2\bd_2}\right)\right|\nonumber\\
&\hspace{2.5 in}\left[\delta\left(\bp-\bp_{+}\right)+\delta\left(\bp-\bp_{-}\right)-
\delta\left(\bp-\bp_{\br_1\bd_1}\right)-\delta\left(\bp-\bp_{\br_2\bd_2}\right)\right].
\end{align}

\end{document}